# Metasurface Polarimeter for Structural Imaging and Tissue Diagnostics

Paul Thrane[1,2]*†, Chao Meng[2]*†, Alexander Bykov[3], Oleksii Sieryi[3], Fei Ding[2], Igor Meglinski[4], Christopher A. Dirdal[1], Sergey I. Bozhevolnyi[2]*

[1]Smart Sensors and Microsystems, SINTEF Digital, Gaustadalleen 23C, 0373 Oslo, Norway.

[2]SDU Centre for Nano Optics, University of Southern Denmark, Campusvej 55, DK-5230 Odense, Denmark.

[3]OPEM, ITEE, University of Oulu, 90014 Oulu, Finland.

[4]College of Engineering and Physical Sciences, Aston University, Birmingham B4 7ET, U.K.

* P.T: paul.thrane@sintef.no; C.M: chao@mci.sdu.dk; S.I.B: seib@mci.sdu.dk

† These authors contributed equally

## Abstract

Histopathology, the study and diagnosis of disease through analysis of tissue samples, is an indispensable part of modern medicine. However, the practice is time consuming and labor intensive, compelling efforts to improve the process and develop new approaches. One perspective technique involves mapping changes in the polarization state of light scattered by the tissue, but the conventional implementation requires bulky polarization optics and is slow. We report the design, fabrication and characterization of a compact metasurface polarimeter operating at 640 nm enabling simultaneous determination of Stokes parameters and degree of polarization with ±2% accuracy. To validate its use for histopathology we map polarization state changes in a tissue phantom mimicking a biopsy with a cancerous inclusion, comparing it to a commercial polarimeter. The results indicate a great potential and suggest several improvements with which we believe metasurface polarimeter based devices will be ready for practical histopathology application in clinical environment.

## Introduction

Polarimetric mapping and characterization of biological tissues, including Mueller matrix and Stokes imaging, have shown significant potential for applications in histological tissue characterization[1–3]. Mueller matrix imaging[4], for example, has been found effective for the detection of morphological and structural alterations associated with cancer[5,6], which has led to the developments of systems such as Mueller matrix imaging endoscopes[7,8]. Similarly, polarization-holographic Mueller matrix methods[9,10] can be applied to the assessment of the 3D morphology of biological tissues with applications in disease diagnosis[11], while Stokes polarimetry has been demonstrated as an effective tool for screening the progression of Alzheimer's disease[12]. These label-free and non-destructive imaging techniques can facilitate the analysis of biological tissue by eliminating the need for conventional sectioning and staining procedures. One such promising technique involves structural imaging of tissue samples by mapping how the state of polarization (SOP) of light is altered by scattering processes in the tissue, and has been applied to the diagnosis and grading of colon cancer[13]. Probing the biological structures in this way not only removes the need to stain the samples and can be done without sectioning and mounting but is also suitable for automated measurement acquisition and analysis. Such a development would significantly speed up diagnosis and subsequent treatment, in addition to reducing intra- and interobserver variability of diagnosis arising from the strong reliance on personal skills and qualifications for conventional methods.

Crucial for such polarization-based measurements are polarimeters – devices that measure the light SOP. Polarimeters fall into one of two categories depending on their principle of operation. Division-of-time systems have a time varying component that measures different properties of the incoming light at different times, determining the full SOP after several measurements. One common way of doing this is by sending the light through a rotating quarter waveplate with a static linear polarizer and then measuring the transmitted intensity, the Fourier decomposition of this signal is then used to find the SOP[14]. Division-of-space systems, on the other hand, split the incoming light into several parts with varying intensities dependent on the SOP, which can then be measured simultaneously and used to calculate the initial SOP[15,16]. Both categories of polarimeters have seen major developments in later years, building on advances in integrated optics to enable compact systems that are faster, more accurate and enable new use cases[17]. Nanostructured surfaces for simultaneous polarimetry have, for example, been used to make polarimeters in optical fibers[18,19] and photonic integrated circuits[20], or by using effects such as surface plasmon polaritons to enhance bandwidth[21]. Various types of measurement strategies have been investigated, such as division-of-focal plane[22], division-of-aperture[23] and in-plane polarimetry[24]. Two-dimensional materials have been used to greatly increase the speed of division-of-time polarimeters by harnessing electronically tunable polarization dependent reflection[25], and even to make photodetectors where the signal is directly dependent on the SOP[26]. Other advancements include spectropolarimeters where the SOP is measured as a function of wavelength[27,28], and polarimetric imaging where the SOP is determined simultaneously for each pixel in an image[29–31].

A common feature among many of these developments is the use of optical metasurfaces (MS), sub-wavelength structured surfaces that can provide independent control over orthogonal light polarizations[32,33]. These MSs are therefore naturally suitable for polarization controlling optics, and the technology has now developed to the point where they are even being included in polarimeter systems for next generation Earth observing satellites[34]. Here, we design and fabricate a thin-film polarimeter based on a plasmonic MS grating that splits the incoming light into several diffraction orders, whose intensities of which are measured and used to calculate the SOP[15] from a single-shot camera acquisition, resulting thereby in rapid polarimetry in a configuration without any moving parts. The MS polarimeter is tailored to be integrated in an experimental setup for digital polarimetric histopathology[35], where the current commercial polarimeter in use limits acquisition time as well as system size and cost. Following calibration, we verify that SOP and degree of polarization (DOP) measurements are sufficiently accurate for this use case. We thereafter benchmark the MS polarimeter against a commercial polarimeter by measuring SOP changes in a tissue phantom designed to mimic the properties of a real biopsy. The results are promising, with the inclusion simulating cancerous tissue being clearly defined in the MS polarimeter measurements. Furthermore, the findings of this study guide the next developments necessary to improve measurement results for real tissue samples and advance the system towards clinical use cases.

## Results

The working principle and overall design of the MS polarimeter are shown in Figure 1. A reflective MS grating is designed to split an incident beam into several different diffraction orders, where the intensity of each order is a function of the polarization state of the incident beam, see Figure 1 a. This is realized by interleaving three separate sub-grating patterns with different periodicities, each designed to separate two orthogonal polarization states (hereafter referred to as the design states) into the +1 and −1 diffraction orders. For example, if the incoming polarization state is equal to one of these design states with DOP of 1, the sub-grating directs all the light into either the +1 or −1 diffraction order according to which of the two states is incoming. On the other hand, if the incoming polarization state is a mix of the two design states, and/or the DOP < 1, the light will be split into both diffraction orders. By measuring the intensity of six such diffraction orders we are able to fully determine the incoming polarization state and DOP, with the six intensities corresponding to the amount of linearly polarized light along 4 axes (0°, 45°, 90° and 135°) as well as left and right-handed circular polarization, as indicated in Figure 1a.

In this work, a reflective gap-surface-plasmon (GSP) MS was used, which has advantages in terms of bandwidth and angular dispersion[36]. Note that the same concept can be used to make a polarimeter using a transmissive MS[16]. While it is possible to design the reflective MS for oblique incident light, but for ease of design and alignment we chose the design presented in Figure 1b, where the input beam is redirected normally onto the MS using a beam splitter. The reflected $0^{th}$ order together with the six diffraction orders go through the beam splitter and are then collimated and focused by a planoconvex lens onto a CMOS sensor, resulting in images of the diffraction orders as shown in Figure 1c. Figures 1d and 1e are images of the experimental prototype, where the MS and planoconvex lens are glued onto opposite sides of the beam splitter which is subsequently mounted on a 6-axis stage for accurate alignment. The 6-axis stage and the CMOS camera are attached to a custom bracket that allows for mounting the polarimeter on an optical cage system. Figure 1f is an image of the diffraction pattern from the MS, where in addition to the $0^{th}$ order and six intentional diffraction orders one observes a row of second order diffraction spots on each side that arise from the superperiodicity of the interleaved MS grating patterns.

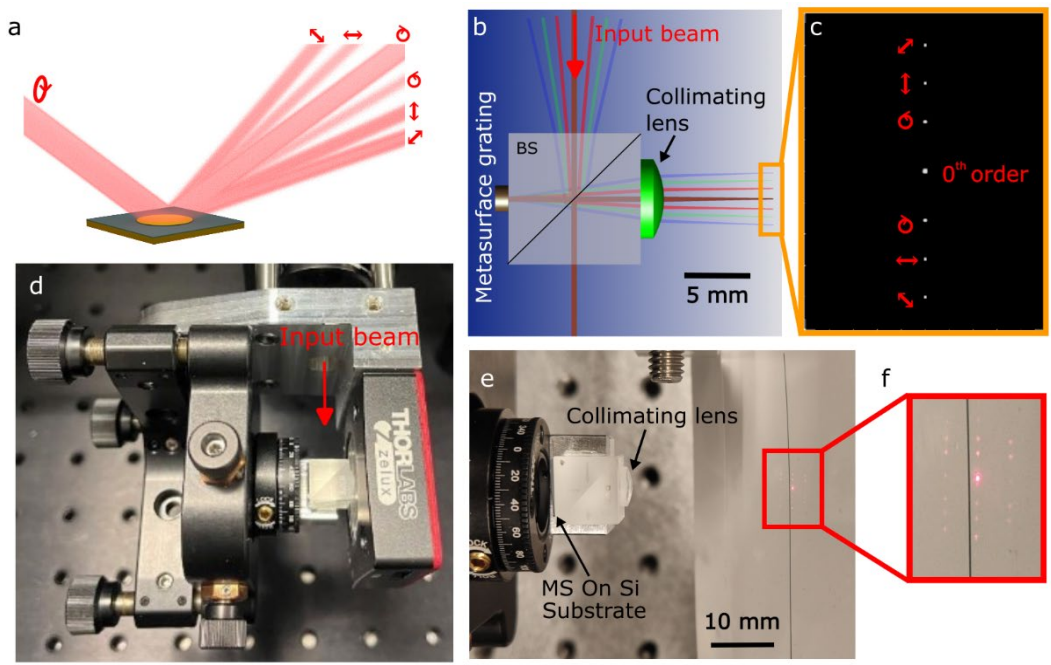

**Figure 1. Concept, optical design and final implementation of the metasurface polarimeter. a** The MS polarimeter concept: a MS splits incoming light with an unknown polarization state into several diffraction orders, each filtering the beam into a separate polarization state as indicated by the red arrows/circles. The original polarization state can then be determined by a single-shot full Stokes measurement from the intensity differences between each of these orders. **b** Design of the MS polarimeter: the input beam enters a beam splitter (BS) and is directed onto the MS grating. The diffraction orders are collimated and focused by a lens onto a CMOS sensor producing the intensity pattern in **c** where red arrows indicate the corresponding polarization states. **d** The assembled MS polarimeter: the MS and a collimating lens are glued onto a BS, which is held by a 6-axis stage used for alignment. The stage is mounted on a bracket for holding a CMOS camera in the correct position and allows for direct attachment to an optical cage system. **e** Close up of the BS on which the MS is mounted on one side, and a collimating lens on the other. The CMOS camera has been replaced by a piece of paper where the diffraction spots are visible – close up in **f**. The $0^{th}$ order is strongest followed by the diffraction orders on either side along the black line. Also visible are higher-order diffraction spots above and below the black line arising from the periodic superlattice of the MS.

The design of the MS grating is described in Figure 2. The basic building block is a rectangular lattice with period $\Lambda_x$ = 250 nm and $\Lambda_y$ = 180 nm, where a gold substrate is covered with a $SiO_2$ spacer of thickness $t_s$ = 50 nm. On top of this dielectric spacer there is a gold nanobrick with side lengths $L_x$, $L_y$ and thickness $t_m$ = 50 nm, see Figure 2a where the direction of the x-, y- and z-axes are also defined. The reflection amplitude and phase of such a unitcell is determined by simulating a periodic lattice of identical nanobricks for each combination of side length values. The results are shown in Figure 2 b and c: the color plots are the simulated reflection amplitudes for light linearly polarized along the *x*- and *y*-directions, with lines indicating the phase shifts for light polarized along the two directions and the difference between them. The values of $t_s$ and $t_m$ have been chosen to get as large phase range as possible while still maintaining a large reflection amplitude for all the relevant geometries. The three sub-gratings are composed of three super-cells denoted by $SC_{xy}$, $SC_{ab}$ and $SC_{rl}$ to respectively split *x/y* linearly polarized, +45°/-45° linearly polarized (hereafter referred to as *a/b*), and right and left circular polarization states (hereafter referred to as *r/l*). For $SC_{xy}$ and $SC_{ab}$ the nanobrick geometries are selected to give an opposite blazed phase to the two orthogonal polarization states as can be seen in Figure 2f). The design of $SC_{rl}$ is based on geometric phase, with the chosen dimensions being the combination of $L_x$, $L_y$ that gives a 180° relative phase shift between *x*-polarized and *y*-polarized light while also having the same reflection amplitude for both polarization states. The phase experienced by circularly polarized light is then opposite for left and right handedness and decided by the angular orientation of the nanobrick, hence the gradual rotation of the structures in $SC_{rl}$ in Figure 2 d give rise to the blazed phase response in Figure 2 f. The simulated reflectance into the various diffraction orders of three standalone gratings composed individually of $SC_{xy}$, $SC_{ab}$ and $SC_{rl}$ is plotted in figures 2 g, 2 h and 2 i. For the design wavelength of λ = 640 nm the efficiency of each is around 40%, with higher efficiencies for longer wavelengths due to the choice of gold for the substrate and nanobricks which results in high absorption for shorter wavelengths. This choice of gold was done on the background of a well-established design and fabrication process, in addition to chemical stability, for shorter wavelengths one might have to use other materials with less absorption such as silver (see Figure S1 in the supplementary information), aluminium or switch to dielectrics.

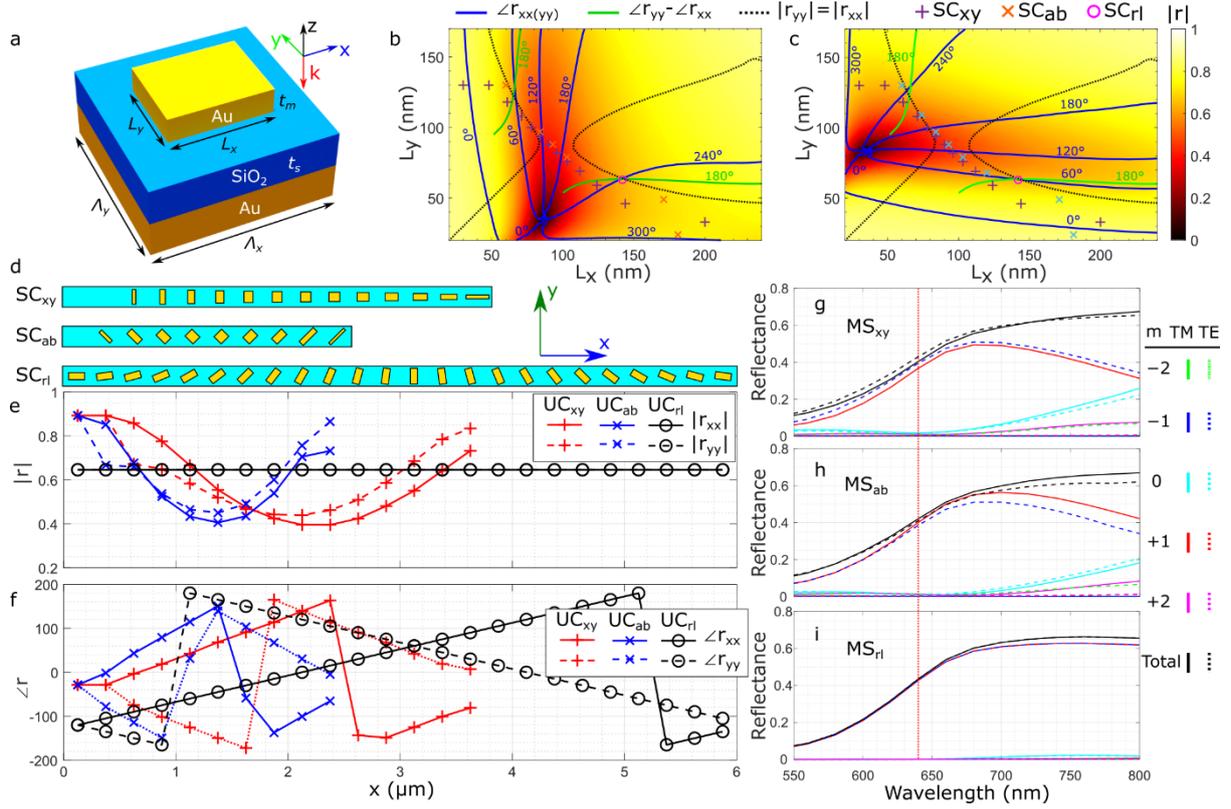

**Figure 2. Design of the MS grating. a** Schematic of a basic unit cell: a gold substrate is covered with $SiO_2$ of thickness $t_s$, on top of which is placed a gold nanobrick with side lengths $L_x$ and $L_y$, and thickness $t_m$. **b** and **c** are the reflection amplitudes (color plots) for light ($\lambda$ = 640 nm) normally incident on a periodic unit cell for different values of $L_x$ and $L_y$. b is for light linearly polarized along the x-axis, likewise for c along the y-axis. Blue lines indicate the phase shift of the reflected light, green lines represent the relative phase shift between light linearly polarized along x- and y-axes, and the black lines mark where these two polarization states have the same reflectivity. Markers (+, x and o) indicate the geometries chosen for the gratings designed to split x/y states (+), a/b states (x), and the single geometry selected for r/l states (o). **d** Illustration of the resulting three different grating supercells that are interleaved to split the beam into six different diffraction orders (in addition to the 0$^{th}$ order). The uppermost is designed to split x/y states, the middle to split a/b states, and the lower to split r/l states. **e** and **f** are reflection amplitudes and phase variations along the length of these gratings. **g-i** Reflectance of several diffraction orders as a function of wavelength for the x/y splitting MS, a/b splitting MS, and the r/l splitting MS, respectively.

The three grating designs have different grating periods to ensure that when interleaved, the final MS splits the different polarization states into well separated diffraction spots. The supercells $SC_{xy}$, $SC_{ab}$ and $SC_{rl}$ have first order diffraction angles of 10°, 15° and 6°, respectively. Several interleaving patterns were investigated, with the chosen one consisting of a repeating pattern with 4 rows of $SC_{xy}$, 4 rows of $SC_{ab}$ and 4 rows of $SC_{rl}$, as seen in Figure 3 a-d. This technique gave better results than switching the grating design for each row, which we attribute to closer resemblance with the periodic simulations used to choose the nanostructures, while still resulting in a high degree of interleaving to avoid measurement errors arising from the beam alignment on the MS. Random interleaving of such sets of identical rows is expected to give similar results while additionally suppressing higher order spots visible in Figure 1 f. The final MS has a circular aperture with diameter 90 μm, and the diffraction patterns when illuminated with 6 different polarization states are shown in Figure 3 e.

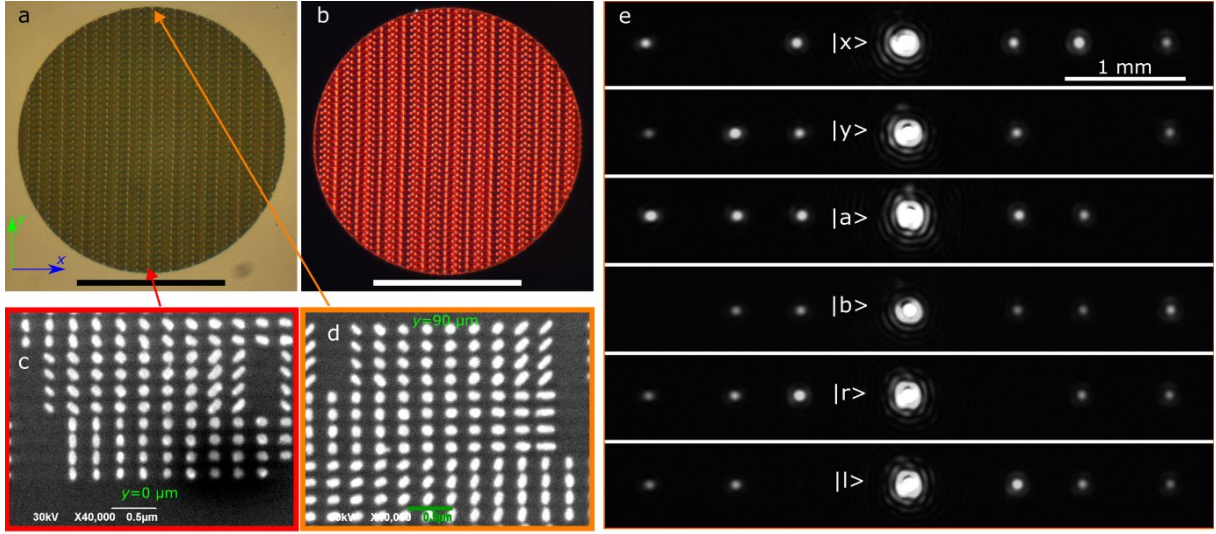

**Figure 3. Fabricated MS polarimeter grating. a** Bright field and **b** dark field microscope images of the fabricated MS polarimeter grating, with scale bars indicating 50 µm. **c** and **d** SEM images from separate sides of the fabricated grating, showing the uniformity of the nanostructures. **e** Diffraction patterns from six different incident polarization states as indicated in each image. The distribution of intensities in different spots is related to their association with light diffraction by specific metasurface grating areas as shown in Figure 1 c. Each of these images individually contain the necessary information to infer the complete Stokes vector of the incoming light.

Ideally, the normalized Stokes parameters of the incoming beam would be calculated directly by

$$S_{\text{uncalibrated}} = \begin{pmatrix} s_1 \\ s_2 \\ s_3 \end{pmatrix} = \begin{pmatrix} (I_x - I_y)/(I_x + I_y) \\ (I_a - I_b)/(I_x + I_y) \\ (I_r - I_l)/(I_x + I_y) \end{pmatrix},$$

with $I_x, I_y, I_a, I_b, I_r$ and $I_l$ being the measured intensities of the six diffraction spots and where we have disregarded the intensity parameter $s_0$. However, to correct for uneven efficiencies of the sub-gratings we include a calibration matrix[37]. The initial approach was to use a 3×3 matrix and calibrate with respect to the Stokes parameters measured by a reference polarimeter. This approach was found to give good results for $s_1$ and $s_3$, but skewed results for $s_2$ due to a difference in diffraction spot efficiency for *a/b* polarization states as seen in Figure 3 e (see the supplementary information for further details). We attribute this to the asymmetric design of SC$_{ab}$, see Figure 2 d, which follows from choosing the geometries based on simulations where the nanobrick and unitcell sides are parallel and subsequently rotating by 45°. In future devices this issue will be solved by simulating rotated structures thus giving a optimal design for SC$_{ab}$.

In this work we correct the asymmetry by using two calibration matrices, one for when $s_2 \leq 0$ and one for $s_2 > 0$, in addition to adding a variable for the DOP. Each of these 4×6 matrices $M$ are determined by measuring the diffraction patterns for a set of *n* known polarization states

$$S = \begin{pmatrix} s_{1,1} & \cdots & s_{1,n} \\ s_{2,1} & \cdots & s_{2,n} \\ s_{3,1} & \cdots & s_{3,n} \\ d_1 & \cdots & d_n \end{pmatrix} = MI = M \begin{pmatrix} I_{x,1}/(I_{x,1}+I_{y,1}) & \cdots & I_{x,n}/(I_{x,n}+I_{y,n}) \\ I_{y,1}/(I_{x,1}+I_{y,1}) & \cdots & I_{y,n}/(I_{x,n}+I_{y,n}) \\ I_{a,1}/(I_{a,1}+I_{b,1}) & \cdots & I_{a,n}/(I_{a,n}+I_{b,n}) \\ I_{b,1}/(I_{a,1}+I_{b,1}) & \cdots & I_{b,n}/(I_{a,n}+I_{b,n}) \\ I_{r,1}/(I_{r,1}+I_{l,1}) & \cdots & I_{r,n}/(I_{r,n}+I_{l,n}) \\ I_{l,1}/(I_{r,1}+I_{l,1}) & \cdots & I_{l,n}/(I_{r,n}+I_{l,n}) \end{pmatrix}.$$

Here $s_{i,j}$ is the $i^{th}$ stokes parameter of the $j^{th}$ polarization state, $d_j$ is the DOP for the $j^{th}$ polarization state, and $I_{i,j}$ is the intensity in diffraction order $i$ of the $j^{th}$ polarization state. The calibration matrix can then be calculated by $\boldsymbol{M} = \boldsymbol{S}\boldsymbol{I}^+$, with $\boldsymbol{I}^+$ being the pseudoinverse of $\boldsymbol{I}$. Any unknown polarization state can then be determined using the measured diffraction intensities and $\boldsymbol{M}$ as

$$S_{\text{calibrated}} = \boldsymbol{M} \begin{pmatrix} I_x/(I_x + I_y) \\ I_y/(I_x + I_y) \\ I_a/(I_a + I_b) \\ I_b/(I_a + I_b) \\ I_r/(I_r + I_l) \\ I_l/(I_r + I_l) \end{pmatrix}.$$

As mentioned, this process is performed twice to form one calibration matrix $\boldsymbol{M}$ for measurements where $I_a \leq I_b$, and another when $I_a > I_b$. After this calibration the three Stokes parameters are renormalized and the value $d$ is used as the DOP setting $d = 1$ if $d > 1$. Figure 4 includes an image of the setup used to calibrate and test the MS polarimeter, together with calibration results for light with DOP = 1 and a short summary of the acquisition procedure: after acquiring an image the average background noise level is subtracted, whereafter the pixel values are summed together for all the pixels in each diffraction order and used to calculate the Stokes parameters. A commercial polarimeter (Thorlabs PAX1000) is used both to calibrate the MS polarimeter and as a reference to assess the measurements. As seen in Figure 4 f and g, the uncalibrated MS polarimeter measurements are more accurate for $s_1$ than for $s_2$ and $s_3$. After calibration the measurements are much more accurate, with most deviations less than ±0.02 for both Stokes parameters and the DOP. Notably, the asymmetry in the $s_2$ measurement disappears after calibration.

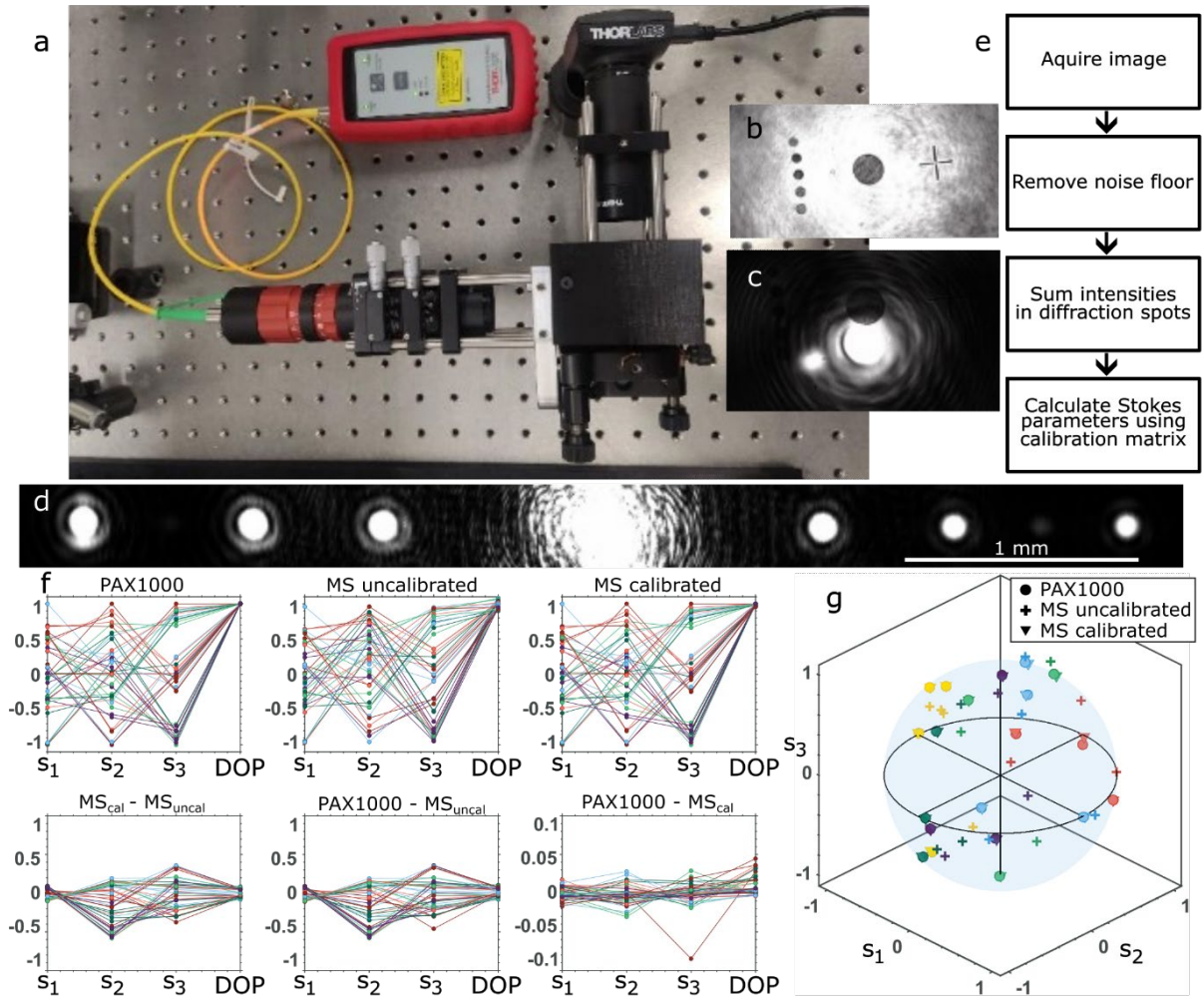

**Figure 4. Characterization of the MS polarimeter with fully polarized light. a** Photo of the calibration setup, where a fiber coupled laser (λ = 640 nm) sends light through polarization optics and into the MS polarimeter. The integrated CMOS camera has been removed in the photo (see Figure 1 d, which makes it possible to check the beam alignment on the MS with a separate camera placed outside the polarimeter (top-right of the picture). **b** and **c** Beam alignment images. The difference between the two images is the width of the incident beam. **d** Diffraction pattern acquisition. After alignment the CMOS camera is inserted into the setup and acquires diffraction pattern images. **e** Acquisition process summary: The average noise value is subtracted from the image, the diffraction spot intensities are summed up and these are used to calculate the Stokes vectors by applying a calibration matrix. **f** Measurement results of different polarization states with DOP = 1. Circles with the same color and connected by a line correspond to the same measurement of a Stokes vector and corresponding DOP. The top row presents measurements done with a Thorlabs PAX1000, the MS polarimeter before and after calibration. The bottom row displays the variation between measurements. **g** Subset of the polarization state measurements plotted on a Poincare sphere. Circles are measured by the Thorlabs PAX1000, + signs are from the MS polarimeter before calibration and triangles are after calibration. Points corresponding to the same measured polarization state are shown in the same color, for most measurements the circles and triangles overlap – indicating good agreement after calibration.

When the DOP is less than 1, there is some light intensity in all the diffraction spots, even for the polarization states that the gratings have been designed for, see example images in Figure 5 a-c. The plots in Figure 5 d-e are DOP measurements of partially polarized states close to the line between $s_2$ = 1 and $s_2$ = -1, where the polarimeter is expected to have the worst performance due to the uneven efficiencies of the $SC_{ab}$ grating. As can be observed in Figure 5 e, for large DOP > 0.8 the difference between the reference polarimeter and the MS polarimeter after calibration stays below 2%, while for lower DOP(i.e., 0.8>DOP>0.2) the difference is below 5%. When the DOP drops below 0.2 the difference becomes much larger, as the intensity variations in the diffraction spots become smaller. For this reason, replacing the CMOS sensor with a counterpart that has better intensity resolution and less noise should improve the accuracy for low DOP measurements.

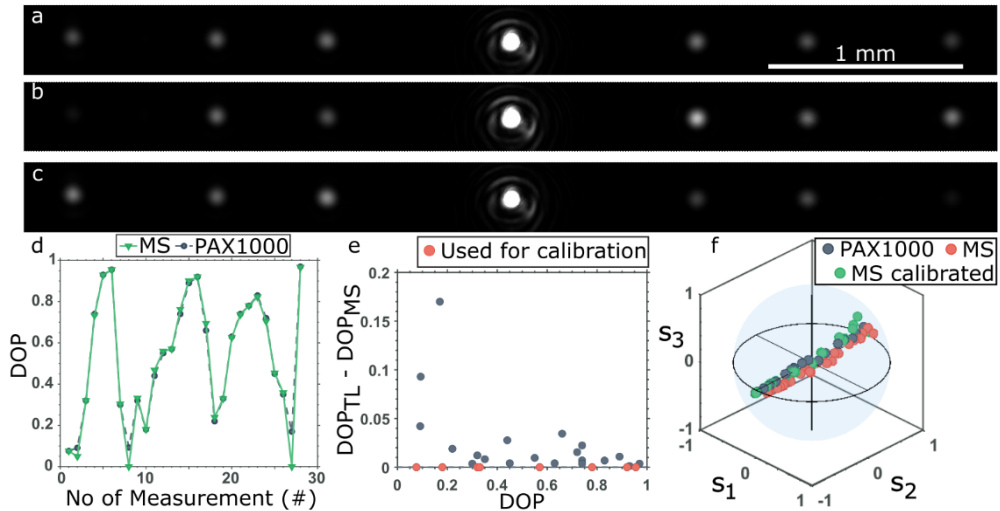

**Figure 5. Characterization of the MS polarimeter with partially polarized light. a-c** Three example images of the diffraction spots when measuring partially polarized light. **d** DOP measurements for a series of measurements where the DOP is cycled between 0 and 1 over four repetitions. **e** Difference between two polarimeters as a function of DOP. **f** Polarization state measurements of partially polarized light in a Poincare sphere. The distance to the origin is proportional to the DOP. Red and green colors respectively indicate the measurements from the MS polarimeter before and after calibration, while blue points are measurements by the reference polarimeter.

The calibration results in Figures 4 and 5 were done using carefully prepared polarization states and simultaneous measurements with both the MS and reference polarimeters. To validate the MS polarimeter in a realistic use case it was tested in a setup for digital histopathology[35]. The calibration matrices used are the same as those in Figure 4, and the results are presented in Figure 6 together with a separate reference measurement done using the commercial polarimeter. Figure 6 a is an image of a tissue phantom consisting of 2 materials made to replicate the geometrical and optical scattering properties of a histological tissue block with a cancerous inclusion[38]. Right-handed circularly polarized light with DOP = 1 is focused onto a spot on this tissue phantom and the scattered light is measured using the MS polarimeter. The spot is scanned across a region of the tissue phantom and Figures 6 b, c are maps showing how the DOP of the scattered light varies across the sample as measured by the two polarimeters. As observed in the DOP maps, both polarimeters give similar results with lower DOP values in regions mimicking the cancerous tissue, although the MS polarimeter exhibits comparatively more noise in these locations. To quantify the differences between the two polarimeters, three areas are selected and marked in blue, green and magenta, respectively corresponding to areas outside the inclusion, where the inclusion is thin and where it is thick – these regions possess different average DOP values in the measurements. A breakdown of all the pixel measurements within each of these areas is given in Figures 6 d-h.

The main source of differences between the two polarimeters can be understood by looking at the intensity distributions measured by the PAX1000 polarimeter shown in Figure 6d. Light scattered from the inclusion areas have on average less intensity and a much higher intensity variation; in the magenta-colored area the intensity is evenly distributed across 7 dBm, while for the blue marked area outside the inclusion most measurements are within 2 dBm of each other. This, combined with the lower DOP and correspondingly lower contrast between diffraction spot intensities for the inclusion areas result in the noticeably larger variation in the measurements from the MS polarimeter for these areas – look for example at the difference in distribution of the magenta-colored measurements in Figures 6 e and 6 f. This difference is still visible when breaking down the polarization states on component basis in Figures 6 g and 6 h, where the MS polarimeter does well for the $s_1$ component but is slightly off for $s_2$ and $s_3$ which have distributions close to zero. Notice also that the MS polarimeter gives a larger DOP than the reference polarimeter for the areas marked with magenta and green. This is attributed to the fact that the calibration matrices were constructed using fully polarized light, and thus are biased towards highly polarized light. Future implementations will mitigate this effect by including also the DOP in the calibration as demonstrated in Figure 5, but for several different polarization states spread evenly across the Poincare sphere rather than just along one line.

In total, the MS polarimeter does a good job characterizing the polarization states and the different regions of the tissue phantom are easily distinguished. However, it was found that for real tissue samples, which have less uniformity than the tissue phantom and thus even greater variability in scattered intensity and DOP, the current implementation of the MS polarimeter is not good enough in terms of dynamic range. While this can be partly mitigated by dynamically adjusting exposure times, the accompanying increase in acquisition time when scanning the whole sample is detrimental for real use cases. A simple solution is to replace the CMOS sensor with a strip array or one photodiode for each diffraction order. This modification would increase the dynamic range and sensitivity, as well as remove the current bottleneck in scanning speed which is limited by a maximum sampling rate of 400 Hz, determined by the speed of the rotating wave plate in the commercial polarimeter. For reference the maximum frame rate of the MS polarimeter is around 600 Hz, limited by the CMOS sensor frame rate for the relevant subset of pixels.

To conclude, we have designed, fabricated and characterized a dedicated MS polarimeter, benchmarking it against a commercial polarimeter for both fully and partially polarized light of wavelength 640 nm. We have found that both design and fabrication MS imperfections can be calibrated away using a simple procedure involving several calibration matrices. Furthermore, we have demonstrated its use for digital histopathology, where it is expected to enable systems for scanning tissue samples that are faster, more compact and cheaper than current laboratory prototypes. Areas of improvement include increase in the dynamic range and mitigation of low DOP measurements, both of which are expected to be straightforwardly implemented in future devices by switching to fewer but more sensitive pixels and improving the set of polarization states used for calibration. We believe that, with these improvements in place, the considered metasurface polarimeter based devices will be ready for practical histopathology applications in clinical environment.

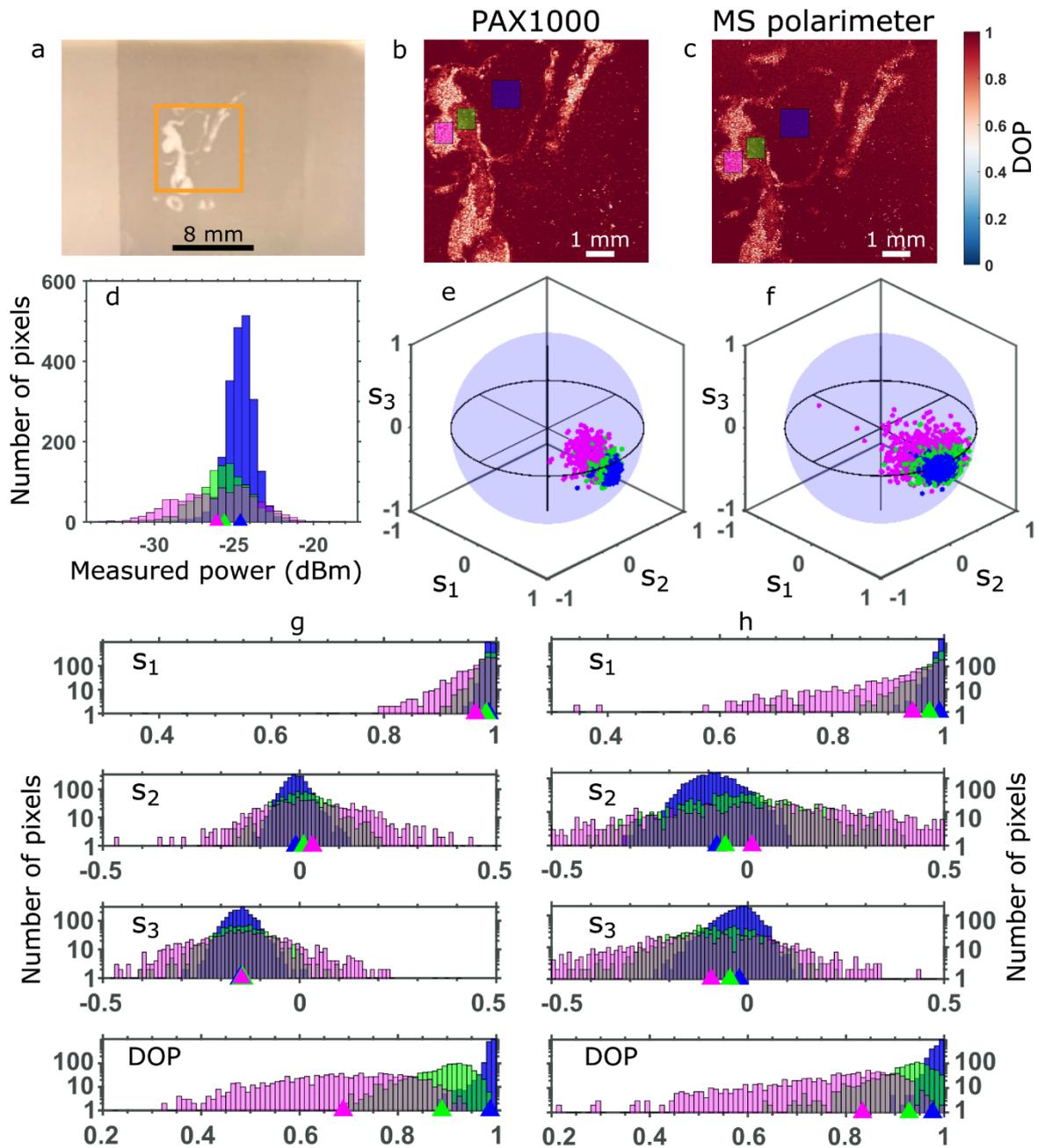

**Figure 6. Benchmarking with tissue phantom. a** Microscope image of a tissue phantom used for comparing the MS polarimeter against a reference polarimeter (Thorlabs PAX1000). Within the area marked with an orange square, the DOP is imaged pointwise using the reference polarimeter **b** and MS polarimeter **c**. The measurement points within the areas marked with blue, green and magenta in **b** and **c** are presented in more detail in the remaining plots **d-h** using the same colors. **d** Histogram of the reflected power as measured by the PAX1000. **e** and **f** Polarization states plotted on a Poincare sphere with the DOP represented as the radius, as measured by the PAX1000, **e,** and the MS polarimeter, **f**. **g** and **h** Histograms of the corresponding Stokes parameters and DOP plotted for the PAX1000, **g,** and the MS polarimeter, **h**. Colored triangles indicate the average values of the various distributions.

# Methods

**Metasurface design and fabrication**

The MS unit cells and super cell gratings were simulated and optimized using the finite element method (FEM) in COMSOL Multiphysics 5.6. The permittivity of Au was based on tabulated values[39] but with a factor 3 increase in the imaginary part of the permittivity for Au material. This adjustment gives closer agreement between simulation and experiments likely due to surface roughness, grain boundary effects and increased damping from a titanium adhesion layer.

The MS was fabricated with electron-beam lithography (EBL) and a lift-off process: Starting with a chip from a polished Si wafer, a 3 nm Ti adhesion layer, a 120 nm Au layer, 3 nm Ti adhesion layer and a 50 nm $SiO_2$ layer were deposited (Tornado 400, Cryofox). This was followed by spin coating of 100 nm PMMA (PMMA A2, MicroChem). The MS pattern was then inscribed using EBL (JEOL JSM-6500F field-emission SEM with a Raith Elphy Quantum lithography system) and subsequent development. The pattern was converted to Au structures by lift-off in acetone after deposition of a 2 nm Ti adhesion layer and 50 nm Au layer (Tornado 400, Cryofox).

**Polarimeter fabrication and calibration**

A model of the MS polarimeter was made in Zemax to optimize the choice of components and their alignment. The polarimeter consists of the MS substrate glued to a non-polarizing beam splitter (Thorlabs BS010) with a plano-convex lens (9 mm focal length) glued on the opposite side as shown in Figure 1, both using UV curing glue (Norland Optical Adhesive 61). The beam splitter was placed on a 6-axis stage connected to a CMOS camera (Thorlabs CS165MU/M) using a custom bracket.

Alignment and calibration were done with the setup shown in Figure 4. A 640 nm fiber coupled laser was connected to a combined collimator/beam expander and sent through polarization controlling optics. Perfect alignment of the beam on the MS was confirmed by imaging the MS surface and beam position through the beam splitter by temporarily removing the integrated CMOS camera. To test the calibration for partially polarized light, a laser was sent through two polarization maintaining fibers, adjusting the relative angle between these two fibers changed the DOP between 0% and 100%.

**Tissue phantom measurements**

A two-component phantom was designed to mimic a histological tissue block, combining adipose and tumorous tissues with complex geometries. A real histological sample of biotissue was used as a basis, with the outlines of the tumorous tissue extracted from a polarimetric scan and converted to a 3D displacement map. The phantom was subsequently 3D printed by stereolithography using UV-curable resins (Formlabs Elastic and Formlabs Clear) with added zinc oxide nanoparticles to model the scattering properties of the adipose tissue. Upon completion, the printed object was rinsed in a solvent bath to remove any remaining resin and then cured in a UV oven to ensure full solidification and structural reinforcement. Finally, the cancerous inclusion of the phantom was added using opaque resin (Photocentric Flexible White and Zortax White Ivory) and the surface was polished after curing in an ultraviolet chamber[38].

The measurements of the tissue phantom were performed in a setup for polarimetric histopathology[35]. A supercontinuum laser was filtered using an acousto-optic tunable filter (640 nm let through) before going through static polarization controlling optics and then being focused onto the sample at an incidence angle of 55°. A 10× objective collected light at an angle of 30°, which was collimated in a 4F system using a 100 μm pinhole before entering the polarimeter. The whole region of interest was mapped pointwise by translating the sample between each measurement. Measurements using the two polarimeters were done separately, thus the mapped areas are slightly different for the two and individual measurement points cannot be compared directly.

## Data availability

The data supporting the findings of this study are available from the corresponding author upon reasonable request.

## Acknowledgements

This research has received funding from the ATTRACT program (European Union Horizon 2020 Research and Innovation Program 101004462); Villum Fonden (37372, 50343, award in Technical and Natural Sciences 2019); Danmarks Frie Forskningsfond (1134-00010B); Norges Forskningsråd (323322); Horizon 2020 CA23125 - The mETamaterial foRmalism approach to recognize cAncer (TETRA). This work was also partially supported by the Research Collaborations grant (1203766815), under the International Science Partnerships Fund funded by the UK Department for Science Innovation and Technology in partnership with the British Council.

## Author contributions

P.T., C.M., C.A.D., S.I.B. and I.M. conceived the idea. MS design and simulation was done by C.M., while P.T. did the system integration design. C.M. fabricated the MS and both P.T. and C.M. did the polarimeter integration. O.S. and A.B. fabricated the tissue phantom. C.M. and O.S. performed the measurements. P.T. and C.M. did the calibration and analyzed the results, with input from S.I.B. and F.D. All authors contributed to the discussion of the results and writing the manuscript. P.T. provided the first draft of the manuscript. C.M., C.A.D. and S.I.B. supervised the work.

## Competing interests

The authors declare no competing interests.

# Supplementary Information: Metasurface Polarimeter for Structural Imaging and Tissue Diagnostics

Paul Thrane[1,2]*†, Chao Meng[2]*†, Alexander Bykov[3], Oleksii Sieryi[3], Fei Ding[2], Igor Meglinski[4], Christopher A. Dirdal[1], Sergey I. Bozhevolnyi[2]*

[1]Smart Sensors and Microsystems, SINTEF Digital, Gaustadalleen 23C, 0373 Oslo, Norway.

[2]SDU Centre for Nano Optics, University of Southern Denmark, Campusvej 55, DK-5230 Odense, Denmark.

[3]OPEM, ITEE, University of Oulu, 90014 Oulu, Finland.

[4]College of Engineering and Physical Sciences, Aston University, Birmingham B4 7ET, U.K.

\* P.T: paul.thrane@sintef.no; C.M: chao@mci.sdu.dk; S.I.B: seib@mci.sdu.dk

† These authors contributed equally

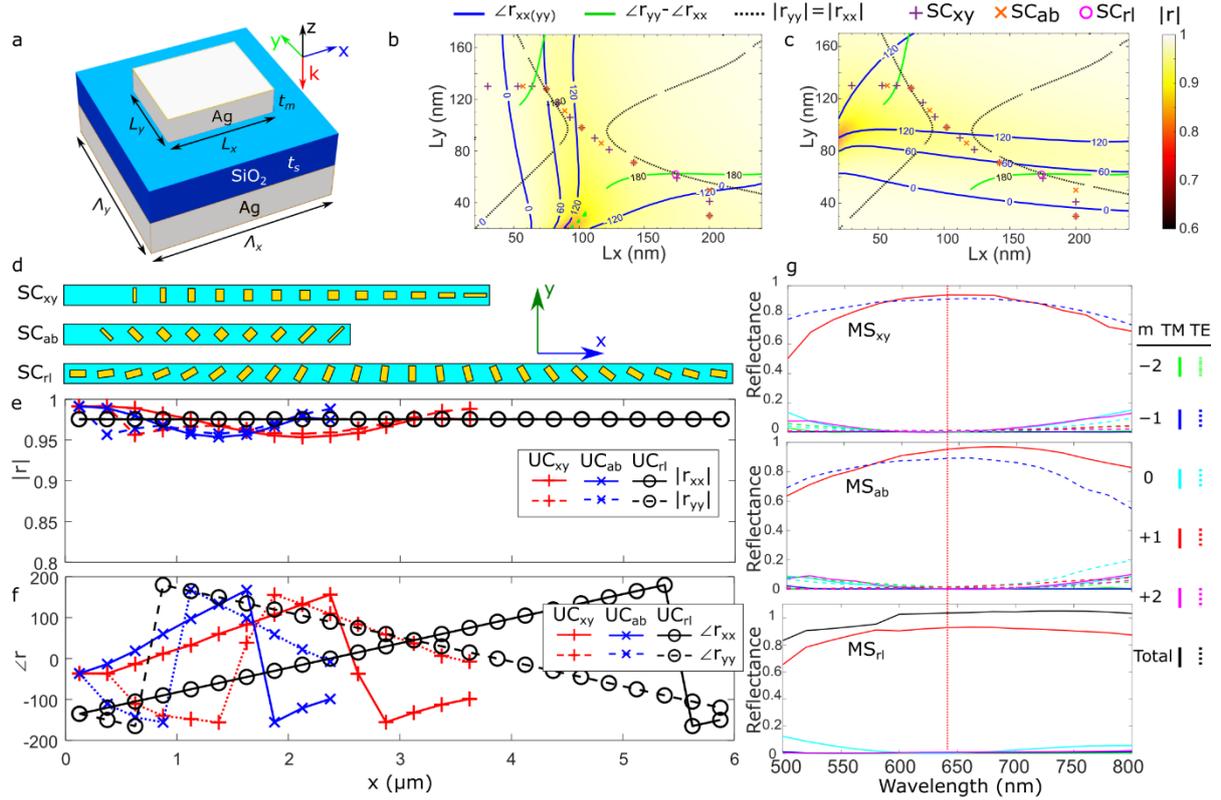

**Figure S1. MS polarimeter grating design based on silver nanostructures.** In the main text, the fabricated MS polarimeter was based on gold nanostructures, but as can be seen in this figure, switching to silver would enable high efficiency polarimetric measurements also for shorter wavelengths. **a** Schematic of a unitcell. **b** and **c** are reflection amplitudes for light linearly polarized light normally incident on a periodic unitcell with nanobrick dimensions $L_x$ and $L_y$, for a wavelength of 640 nm. **d** is a drawing of the periodic cells used to make the three supercells. **e** and **f** are the reflection amplitudes and phases for the chosen geometries, while the plots in **g** are the simulated diffraction order efficiencies for the three diffraction gratings before being interleaved.

# Supplementary Note

In the main text we present a method of using two 4×6 matrices to do the MS polarimeter calibration, which was found to give the best results. Here we present the initial procedure we investigated which uses a single 3x6 calibration matrix $M$, determined by measuring the diffraction patterns for a set of *n* known polarization states

$$S = \begin{pmatrix} s_{1,1} & \cdots & s_{1,n} \\ s_{2,1} & \cdots & s_{2,n} \\ s_{3,1} & \cdots & s_{3,n} \end{pmatrix} = MI = M \begin{pmatrix} I_{x,1}/(I_{x,1}+I_{y,1}) & \cdots & I_{x,n}/(I_{x,n}+I_{y,n}) \\ I_{y,1}/(I_{x,1}+I_{y,1}) & \cdots & I_{y,n}/(I_{x,n}+I_{y,n}) \\ I_{a,1}/(I_{a,1}+I_{b,1}) & \cdots & I_{a,n}/(I_{a,n}+I_{b,n}) \\ I_{b,1}/(I_{a,1}+I_{b,1}) & \cdots & I_{b,n}/(I_{a,n}+I_{b,n}) \\ I_{r,1}/(I_{r,1}+I_{l,1}) & \cdots & I_{r,n}/(I_{r,n}+I_{l,n}) \\ I_{l,1}/(I_{r,1}+I_{l,1}) & \cdots & I_{l,n}/(I_{r,n}+I_{l,n}) \end{pmatrix}.$$

Here $s_{i,j}$ is the *i*[th] stokes parameter of the *j*[th] polarization state, and $I_{i,j}$ is the intensity in diffraction order *i* of the *j*[th] polarization state. The calibration matrix can then be calculated by $M = SI^+$, with $I^+$ being the pseudoinverse of $I$. Any unknown polarization state can then be determined using the measured diffraction intensities and $M$ as

$$S_\text{calibrated} = M \begin{pmatrix} I_x/(I_x+I_y) \\ I_y/(I_x+I_y) \\ I_a/(I_a+I_b) \\ I_b/(I_a+I_b) \\ I_r/(I_r+I_l) \\ I_l/(I_r+I_l) \end{pmatrix}.$$

The results are shown in Supplementary Figures S2-S4.

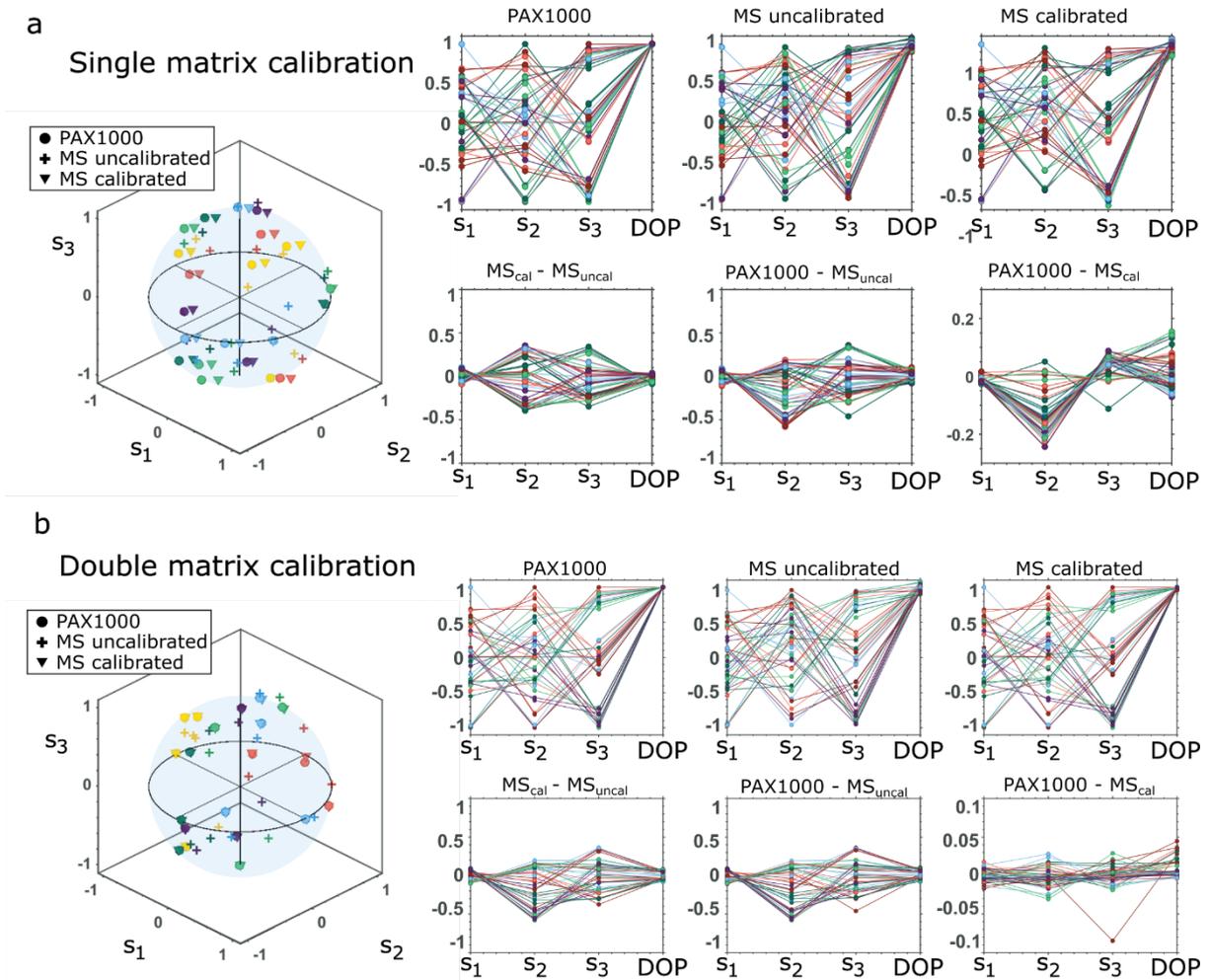

**Figure S2. Characterization of the MS polarimeter with fully polarized light. a** Calibration results when using a single calibration matrix. **b** Calibration results shown in Figure 4 in the main text, done using two calibration matrices. When comparing the two plots labeled "PAX1000 – $MS_{cal}$" it is obvious that the $S_2$ measurements are skewed when using a single matrix. This is due to an asymmetric design of $SC_{ab}$, which is why we chose to use two separate matrices for when $S_2 \geq 0$ and $S_2 < 0$.

## a Single matrix calibration

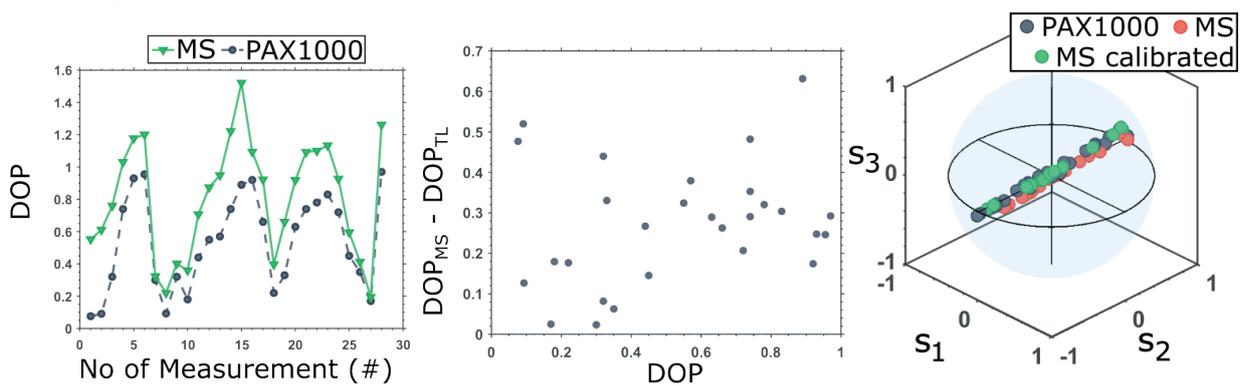

## b Double matrix calibration

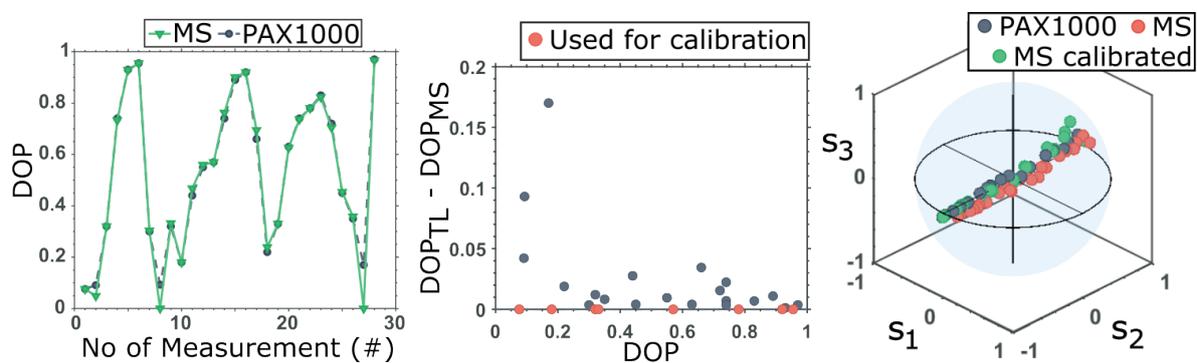

**Figure S3. Characterization of the MS polarimeter with partially polarized light. a** Calibration results when using a single calibration matrix. **b** Calibration results when using two matrices and adding a row in the calibration matrices for adjusting the DOP, as presented in Figure 5 of the main text. Notice that the DOP is clearly wrong for the single matrix method, with the DOP > 1 for several measurements calculated as the norm of the three Stokes parameters. In a all the measurement points have been included in the calibration, while for the plots in b only a subset have been used – as indicated by red dots in the middle plot.

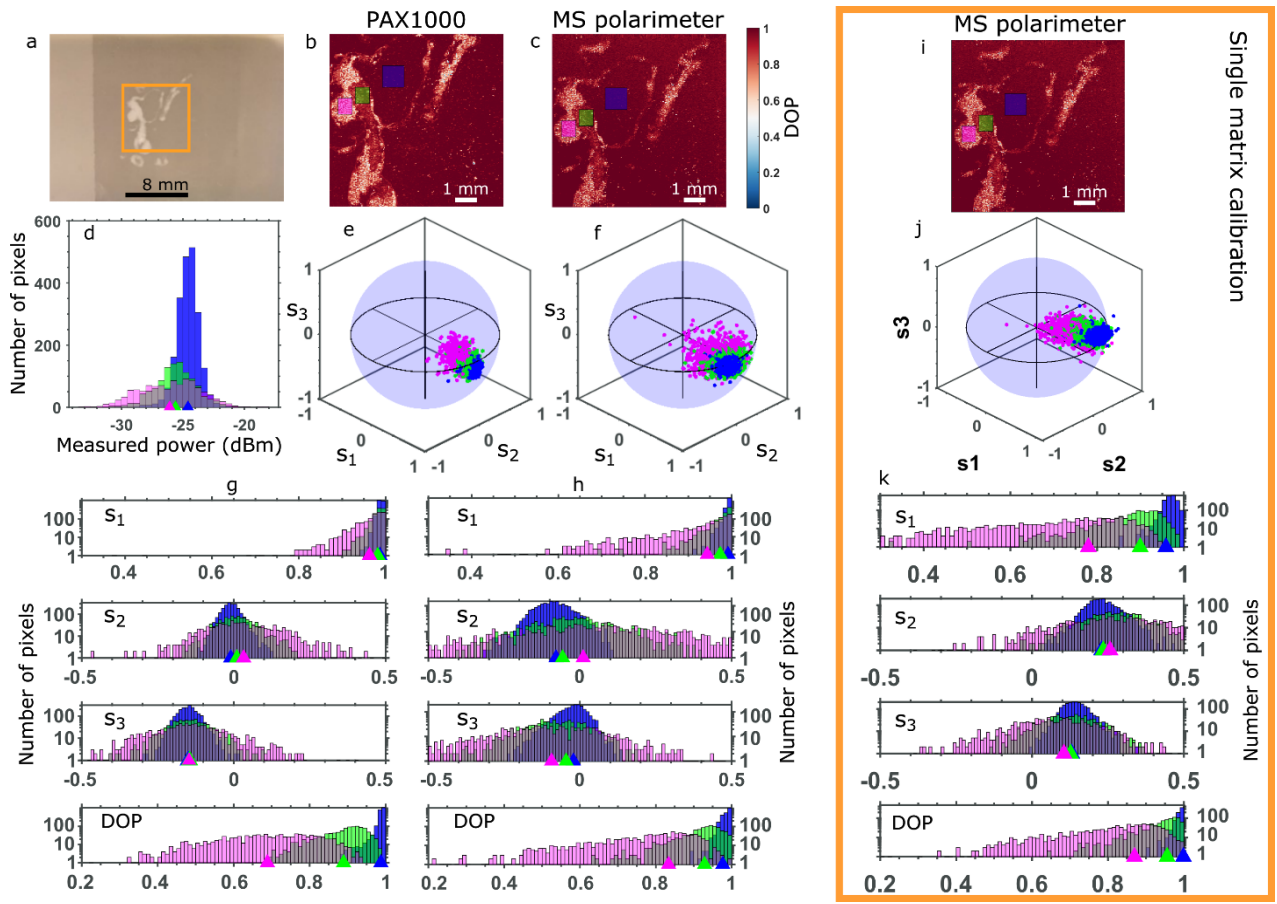

**Figure S4. Benchmarking with tissue phantom. a-h** Benchmarking results presented in Figure 6 of the main text. **i-k** Benchmarking results when using a single calibration matrix. As mentioned in Figure S2, when using a single calibration matrix the $S_2$ measurement is skewed which can be seen as a shift in $S_2$ towards positive values. Including the calibration for the DOP also improves the $S_1$ and $S_3$ measurements, even though the calibration matrices were constructed using only polarization states with DOP = 1. Adding partially polarized light to the calibration procedure is expected to improve the results.